\documentclass[12pt]{article}
\pdfoutput=1
\usepackage{amsmath, amssymb}
\usepackage{graphicx}

\begin{document}

\begin{titlepage}
\begin{center}

\hfill ICRR-report-605-2011-22 \\
\hfill IPMU11-0016 \\

\vspace{2.5cm}
{\Large\bf 
Pure Gravity Mediation with $m_{3/2} = 10$--$100$\,TeV
}
\vspace{2.5cm}

{\bf Masahiro Ibe}$^{(a,b)}$,
{\bf Shigeki Matsumoto}$^{(b)}$,
and
{\bf Tsutomu T. Yanagida}$^{(b)}$

\vspace{1.5cm}
{\it
$^{(a)}${\it ICRR, University of Tokyo, Kashiwa, 277-8582, Japan} \\
$^{(b)}${\it IPMU, TODIAS, University of Tokyo, Kashiwa, 277-8583, Japan} \\
}
\vspace{3.0cm}

\abstract{
Recently, the ATLAS and CMS collaborations reported exciting hints of a Standard Model-like Higgs boson with a mass around $125$\,GeV.
Such a Higgs boson mass can be easily obtained in the minimal supersymmetric 
Standard Model based on 
the ``pure gravity mediation model" 
where the sfermion masses and the Higgs mass parameters are in tens to hundreds TeV
range while the gauginos are in the hundreds GeV to TeV range.
In this paper, we discuss detalis of the gaugino mass spectrum
in the pure gravity mediation model.
We also discuss the signals of the model at the current and future experiments
such as cosmic ray observations and the LHC experiments.
In particular, we show that the parameter space which is consistent 
with the thermal leptogenesis can be fully surveyed experimentally 
in the foreseeable future.
}

\end{center}
\end{titlepage}
\setcounter{footnote}{0}

\section{Introduction}
The pure gravity mediation model investigated in Ref.\,\cite{Ibe:2011aa} 
is a surprisingly simple model of 
the supersymmetric Standard Model (SSM). 
There, the scalar bosons obtain supersymmetry (SUSY)
breaking masses from a SUSY breaking sector via tree-level interactions 
in supergravity\,\cite{Nilles:1983ge}.
The Higgs mixing mass parameters,  $\mu$-term
and $B$-term, are also generated via tree-level interactions of supergravity\,\cite{Inoue:1991rk}.
Due to the tree-level mediation, the scalar boson masses
and the Higgs mixing mass parameters are expected to be of the order of the gravitino
mass, $m_{3/2}$.
The gaugino masses are, on the other hand, generated at the one-loop 
level\,\cite{Giudice:1998xp,Randall:1998uk,hep-ph/9205227}. 
Thus, the pure gravity mediation model predicts a hierarchical spectrum.
The greatest benefit of the pure gravity mediation is that 
the model requires no additional fields to realize the above spectrum. 
Therefore, the pure gravity mediation model is the bare-bones model
of the supersymmetric Standard Model.

The pure gravity mediation model is particularly successful when the gravitino mass 
is in the range of  $m_{3/2} =10$--$100$\,TeV.
The first advantage is the alleviation of the cosmological gravitino problem\,\cite{Pagels:1981ke,kkm}.
Especially, the model does not suffer from the gravitino problem even for a 
very high reheating temperature
after inflation, $T_R \gtrsim \sqrt{3}\times 10^9$\,GeV, 
which is essential for the successful thermal leptogenesis\,\cite{leptogenesis}.
The second advantage is that the model has a good candidate for dark matter.
For the above  gravitino mass range, the lightest superparticle (LSP) which is 
neutral wino in the pure gravity mediation obtains a mass in hundreds GeV to TeV range.
The neutral wino in this mass range is a good candidate of weakly interacting particle 
dark matter\,\cite{Gherghetta:1999sw,hep-ph/9906527}.
Moreover, as emphasized in Ref.\,\cite{Ibe:2004tg,Ibe:2011aa}, 
the relic density of the neutral wino can be consistent with the observed value 
when we assume the thermal leptogenesis.
Therefore, the pure gravity mediation model goes quite well with the thermal leptogenesis.
Another but an important advantage in cosmology is  that 
the model does not suffer from the cosmological Polonyi problem\,\cite{Polonyi} 
since no singlet SUSY breaking fields are required in the model.%
\footnote{See also Ref.\,\cite{hep-ph/0605252} for the Polonyi problem 
in dynamical supersymmetry breaking models.} 
In addition to those advantages in cosmology,
the problems of flavor-changing neutral currents and CP violation in the SSM
are highly ameliorated thanks to the large masses for squarks and sleptons. 
The unification of the gauge coupling constants at the very high energy scale also provides a strong motivation to the model.%
\footnote{In fact, the gauge coupling constants unify at around $10^{16}$\,GeV at a few percent level
even for a rather large $\mu$-term of $10$--$100$\,TeV.
It should be noted that the scale of the coupling unification is slightly lower than
the conventional SSM for $m_{3/2}=10$--$100$TeV about a factor of two or so. 
Thus, the model predicts a slightly shorter proton lifetime via the so-called dimension six operators,
$\tau_p \lesssim 10^{35}$\,yrs, which is within the reach of the Hyper-Kamiokande 
Experiment\,\cite{Abe:2011ts}.
}

In Ref.\,\cite{Ibe:2011aa},  two of the authors (M.I. and T.T.Y) discussed 
the lightest Higgs boson mass of the minimal SSM (MSSM) based on
 the pure gravity mediation model.
There, we showed that the lightest Higgs boson mass is required to be below about 
$128$\,GeV if we assume the thermal leptogenesis.
This requirement has been shown  to be consistent with 
the most recent experimental constraints on the Higgs boson mass, $m_h>115.5$\,GeV and 
 $m_h <127$\,GeV at 95\,\%C.L. reported by ATLAS\,\cite{ATLAS} and CMS collaborations\,\cite{CMS}.
Furthermore, as shown in Ref.\,\cite{Ibe:2011aa}, 
the pure gravity mediation model can easily explain 
the rather heavy Higgs boson mass around $125$\,GeV 
which is tantalizingly hinted by  ATLAS and CMS collaborations.

In this letter, we discuss phenomenological, cosmological and astrophysical 
aspects of the pure gravity mediation model.
In particular, in this paper, we concentrate ourselves on the 
the parameter space of the model which is consistent with the thermal leptogenesis.
As we will show, such a parameter space can be fully tested by the  observation 
of the cosmic rays, especially by the observation of the anti-proton flux
in the foreseeable future.
We also discuss the strategies of the discoveries and the measurements
of the gauginos at the Large Hadron Collider (LHC) experiments.
There,  the distinctive gaugino mass spectrum in the pure gravity mediation
model plays important roles.

The organization of the paper is as follows.
In section\,\ref{sec:pSUGRA}, we review the model with pure gravity mediation.
There, we discuss the details of the gaugino spectrum.
In section\,\ref{sec:signals}, we discuss the phenomenological, cosmological
and astrophysical aspects on the model.
The final section is devoted to our conclusions.

\section{Pure gravity mediation model}\label{sec:pSUGRA}
\subsection{Mass spectrum}
In the pure gravity mediation model, 
the only new ingredient other than the MSSM fields
is a (dynamical) SUSY breaking sector.
There, the scalar bosons obtain the soft SUSY
breaking squared masses mediated by tree-level interactions in supergravity.
With a generic K\"ahler potential, all the soft squared masses of the scalar bosons
are expected to be of the order of the gravitino mass\,\cite{Nilles:1983ge}.  
The soft SUSY breaking scalar trilinear coupling, the $A$-terms, are, on the other hand,  
expected to be suppressed in supergravity at the tree-level.

In the pure gravity mediation model, 
the Higgs mixing $\mu$ and $B$ parameters can be also generated 
via tree-level interactions in supergravity.
In fact, if the Higgs doublets are not charged under any special symmetries,  
we expect the following K\"ahler potential,
\begin{eqnarray}
    K \ni c H_{u}H_{d} + 
    \frac{c'}{M_{PL}^{2}} X^{\dagger} X H_{u}H_{d} + h.c..
\end{eqnarray}
Here, $X$ denotes a chiral SUSY breaking field
in a (dynamical) SUSY breaking sector, 
$M_{PL}$ is the reduced Planck scale, 
and $c$ and $c'$ are coefficients of ${ O}(1)$.
Through the above K\"ahler potential, the $\mu$- and the
$B$-parameters\,\cite{Inoue:1991rk,Ibe:2006de}
\begin{eqnarray}
    \label{eq:Muterm}
    \mu_H &=& c m_{3/2},\\
    \label{eq:Bterm}
    B \mu_H &=& c m_{3/2}^{2} + c'\frac{|F_X|^2}{M_{PL}^{2}},
\end{eqnarray}
where $F_X$ is the vacuum expectation value of the $F$-component of $X$.
Therefore, the  $\mu$- and $B$ Higgs mixing parameters are also expected to be 
of ${ O}(m_{3/2})$.%
\footnote{If the SUSY breaking sector has a singlet Polonyi field,
the so-called Giudice-Masiero mechanism\,\cite{Giudice:1988yz} 
can also generate the $\mu$  and $B$ Higgs mixing parameters of $O(m_{3/2})$.
In that case, however, the model suffers from the Polonyi problem\,\cite{Polonyi}.
}

For the gaugino masses, on the other hand, tree-level contributions 
in the supergravity are extremely suppressed since we have no  
SUSY breaking fields which are singlet under any symmetries.
At the one-loop level, however, the gaugino masses 
are generated  without having singlet SUSY breaking fields, which is  
called the anomaly mediated contributions\,\cite{Giudice:1998xp,Randall:1998uk}.
Besides, the gauginos also obtain contributions from the heavy Higgsino
threshold effects at the one-loop level.
Putting these one-loop contributions together, 
the gaugino masses at the energy scale of the scalar boson masses, 
$M_{\rm SUSY} = O(m_{3/2})$, are given by\,\cite{Giudice:1998xp,Gherghetta:1999sw}
\begin{eqnarray}
    M_{1} &=& 
    \frac{33}{5} \frac{g_{1}^{2}}{16 \pi^{2}}\left(m_{3/2}+ \frac{1}{11} L\right)\ ,
    \label{eq:M1} \\
    M_{2} &=& 
    \frac{g_{2}^{2}}{16 \pi^{2}} \left(m_{3/2} + L\right)\ ,
        \label{eq:M2}     \\
    M_{3} &=&  -3 \frac{g_3^2}{16\pi^2} m_{3/2}\ .
    \label{eq:M3}
\end{eqnarray}
Here, the subscripts $M_a$, $(a=1,2,3)$ correspond to the gauge groups of the Standard Model
$U(1)_Y$, $SU(2)_L$ and $SU(3)$, respectively. 
In the above expressions, the terms proportional to $m_{3/2}$ denote the anomaly mediated 
contributions and the terms proportional to $L$ denote the Higgsino threshold
contributions.
The parameter $L$ is given by
\begin{eqnarray}
    L \equiv \mu_H \sin2\beta 
    \frac{m_{A}^{2}}{|\mu_H|^{2}-m_{A}^{2}} \ln \frac{|\mu_H|^{2}}{m_{A}^{2}} 
        \label{eq:L}\ ,
\end{eqnarray} 
where $m_A$ denotes the mass of the heavy Higgs bosons,
and $\tan \beta$ is the ratio of the vacuum expectation values of 
the up-type Higgs boson $H_u$ and the down-type Higgs boson $H_d$.
As we will see in the next subsection, the size of $L$ is expected to be of the order of the gravitino 
mass in the pure gravity mediation model\,\cite{Ibe:2011aa}. 
Therefore, the wino mass obtains comparable contributions 
from the anomaly mediated effects and the Higgsino threshold effects.
This facts have a great impacts on the testability of the pure gravity mediation model 
at the LHC experiments.

Before closing this section, we should emphasize the difference
of the pure gravity mediation model from the 
the Split Supersymmetry\,\cite{hep-th/0405159,hep-ph/0406088,hep-ph/0409232}.
In the first place, the Split Supersymmetry mainly considers
a scalar mass scales much higher than that in the pure gravity mediation model, 
i.e. $M_{\rm SUSY} \gg 10^{4-6}$\,GeV.
Thus, the anomaly-mediated gaugino masses should be suppressed in the
Split Supersymmetry, while we rely on the anomaly-mediated gaugino masses 
in the pure gravity mediation model.%
\footnote{See discussions on the possible cancellation of the 
anomaly-mediated gaugino masses\,\cite{hep-ph/0409232,Izawa:2010ym}.
}
Thus, the pure gravity mediation model is more close to
the PeV-scale Supersymmetry\,\cite{hep-ph/0411041}
and the Spread Supersymmetry\,\cite{arXiv:1111.4519}.
Another important and more practical difference is the size of $\mu$-term.
In the Split Supersymmetry, it is assumed that the higgsinos are also in the TeV range.
Thus, the absence of the Higgsino in the TeV range will be a crucial observation
to distinguish the pure gravity mediation model from the Split Supersymmetry.
Furthermore, as we will see below, such a large $\mu$-term leads to a peculiar gaugino spectrum
in the pure gravity mediation model.
Thus, we can also distinguish these models by carefully examining the gaugino mass spectrum.

\subsection{Details on gaugino masses}

\begin{figure}[t]
\begin{center}
\begin{minipage}{.325\linewidth}
  \includegraphics[width=\linewidth]{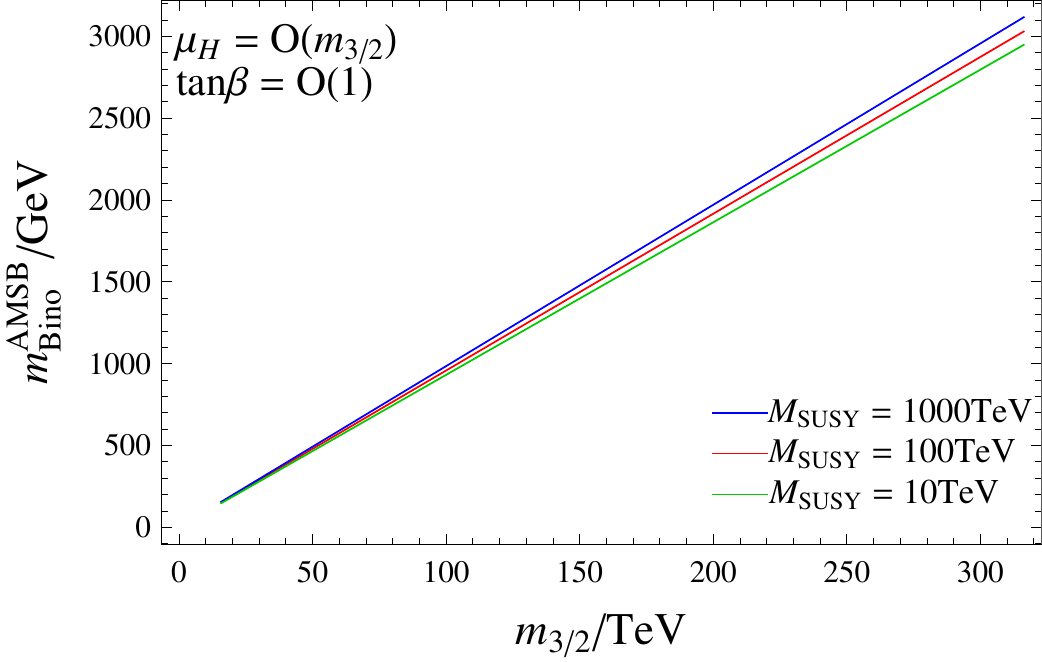}
\end{minipage}
\begin{minipage}{.325\linewidth}
  \includegraphics[width=1\linewidth]{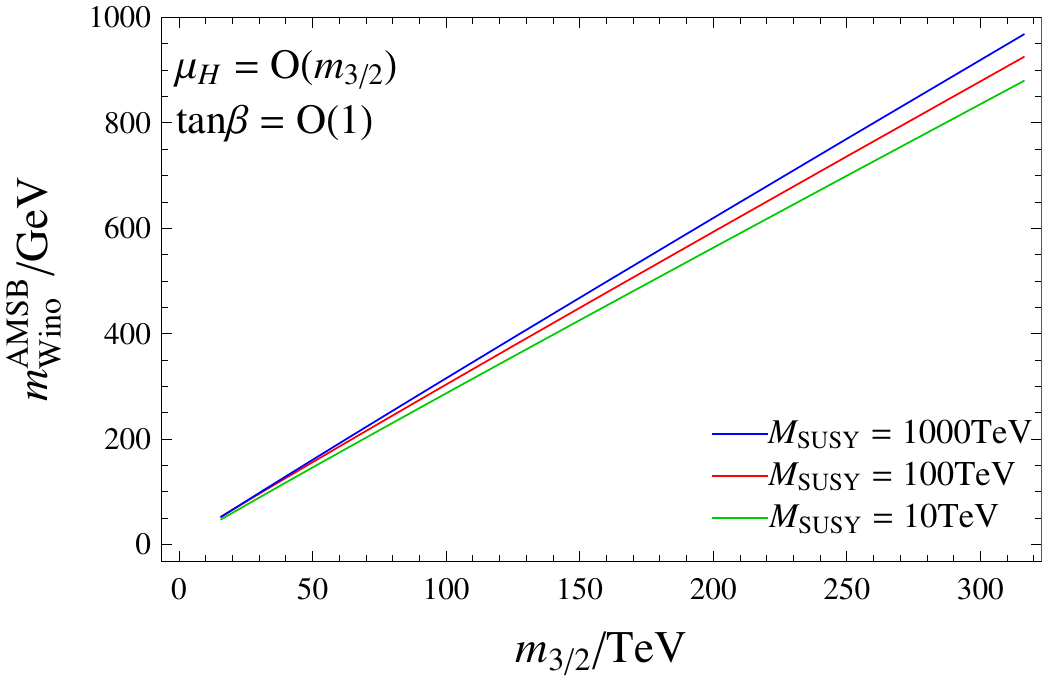}
  \end{minipage}
  \begin{minipage}{.325\linewidth}
  \includegraphics[width=.93\linewidth]{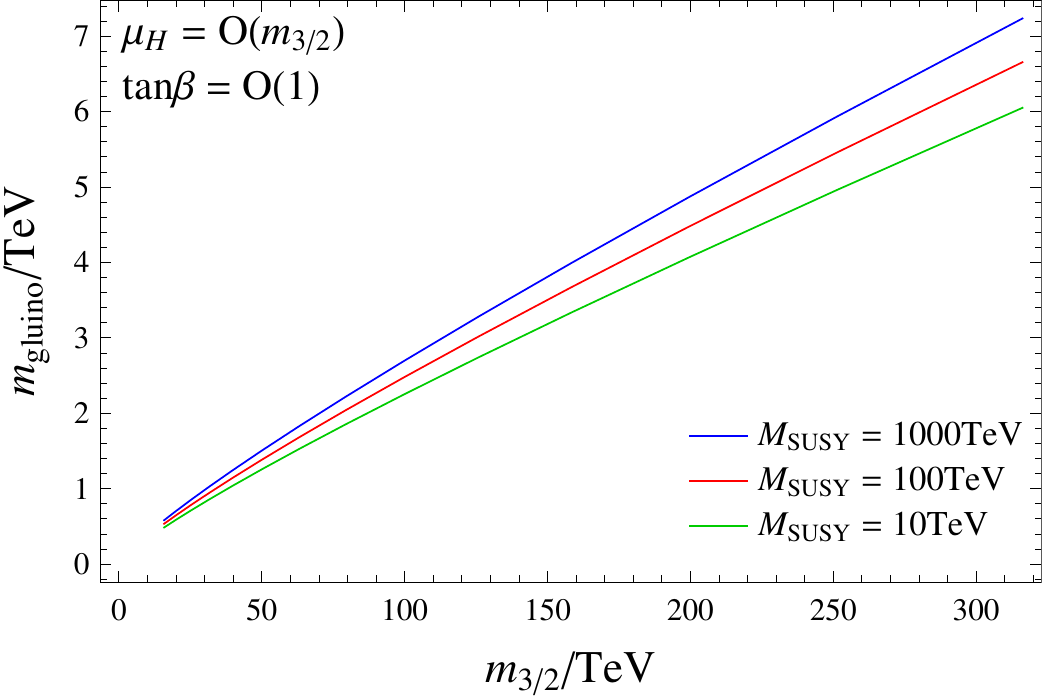}
  \end{minipage}
\caption{\sl \small
The anomaly mediated contributions to the gaugino masses (denoted by AMSB).
The each line corresponds to the heavy scalar threshold scale
$M_{\rm SUSY} = 10,100$ and $1000$\,TeV from bottom to up.
In the figure, we have taken $\mu_H = O(m_{3/2})$ and $\tan\beta = O(1)$,
although they are not sensitive to those parameters.
}
\label{fig:gaugino}
\end{center}
\end{figure}

As discussed above, the pure gravity mediation model predicts
that the sfermions, Higgsinos and the heavier Higgs bosons in the MSSM have 
masses of the order of the gravitino mass, $m_{3/2}=10$--$100$\,TeV. 
Therefore, the only accessible particles at the collider experiments 
in the foreseeable future are the gauginos.
In this subsection, we give detailed analysis on the  gaugino mass
spectrum in the pure gravity mediation model.

First, let us consider the anomaly mediated contributions to the gaugino masses.
As we see from Eqs.(\ref{eq:M1})-(\ref{eq:M3}), the wino is the lightest gauginos
for $L = 0$.
This feature is related to the fact that the $SU(2)_L$ gauge coupling constant 
is the least scale dependent out of the three gauge coupling constants.
In Fig.\,\ref{fig:gaugino}, we show the anomaly mediated gaugino masses
as a function of the gravitino mass.
The figure shows that the gaugino masses are roughly given by,
\begin{eqnarray}
 m_{\rm bino} &\simeq& 10^{-2}m_{3/2}\ ,\\
  m_{\rm wino} &\simeq& 3\times 10^{-3}m_{3/2}\ ,\\
   m_{\rm gluino} &\simeq& (2-3)\times 10^{-2} \, m_{3/2} \ .
\end{eqnarray}
with small dependences on the heavy scalar threshold scale, $M_{\rm SUSY}$.
Thus, for the wino mass $m_{\rm wino} = 300\,$GeV, for example,
the gluino mass is heavier than $2$\,TeV if the anomaly mediated contributions 
dominate the gaugino masses.

Now, let us estimate the typical size of $L$ in the pure gravity mediation model
which parametrize the Higgsino threshold contributions to the gaugino masses.
Let us remember that we require
one of the linear combinations of the two Higgs bosons,
$ h = \sin\beta H_{u} - \cos\beta H_{d}^{*}$ remains 
very light for successful electroweak symmetry breaking. 
In terms of the Higgs mass parameters, the above fine-tuning 
condition requires,  
\begin{eqnarray}
\label{eq:tuning}
   (|\mu_H|^2+m_{H_u}^2)    (|\mu_H|^2+m_{H_d}^2) - (B\mu_H)^2 \simeq 0\ ,
\end{eqnarray}
while the Higgs mixing angle is related to the Higgs mass parameters by,
\begin{eqnarray}
\label{eq:angle}
  \sin2\beta = \frac{2 B\mu_H}{m_A^2} \ ,  \quad (m_A^2 = m_{H_u}^2 + m_{H_d}^2 + 2 |\mu_H|^2)\ .
\end{eqnarray}
Here, $m_{H_{u,d}}^2$  denote the soft SUSY breaking squared masses 
of the two Higgs doublets, $H_u$ and $H_d$.
These conditions show that the mixing angle $\beta$ 
is expected to be of ${O}(1)$,
since all the mass parameters of the Higgs sector (except for a fine-tuning condition) 
are of the order of the gravitino mass in the pure gravity mediation model.%
\footnote{Hereafter, we use a phase convention 
where  $B\mu_H$ is real and positive.
} 

\begin{figure}[t]
\begin{center}
  \includegraphics[width=.5\linewidth]{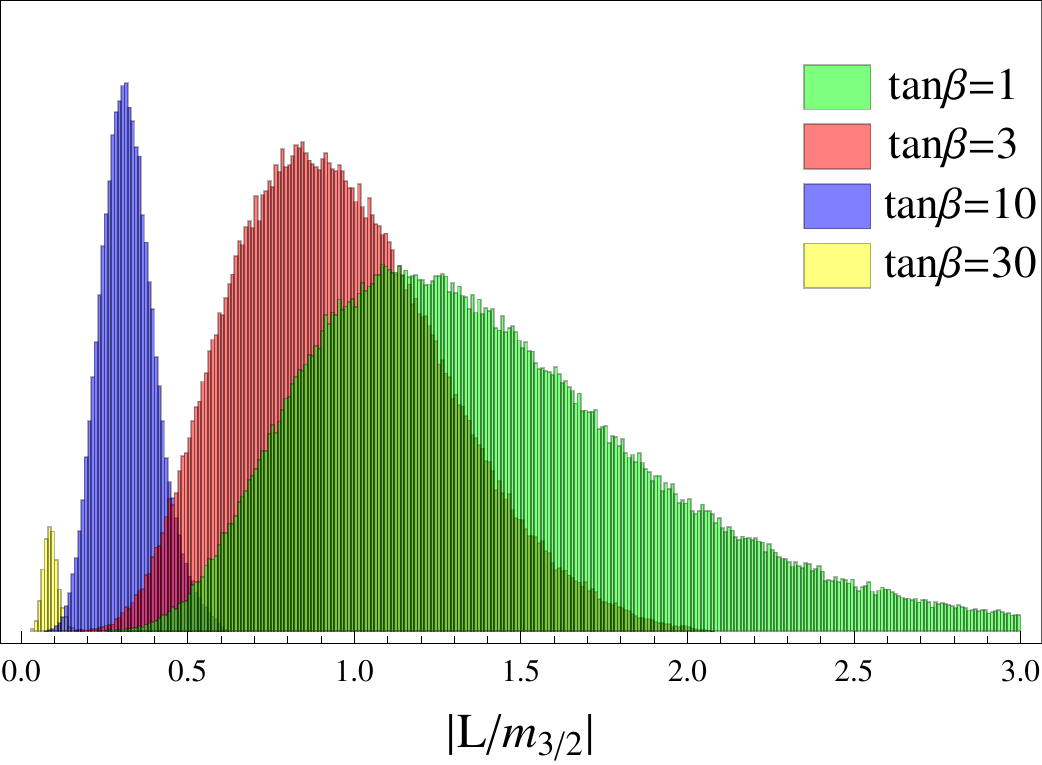}
\caption{\sl \small
The typical values of $|L/m_{3/2}|$ for $\tan \beta = 1,3, 10$ and $30$.
The unit of the vertical axis is arbitrary.
We have distributed $\mu_H$ and $B$ from $m_{3/2}/3$ to $3m_{3/2}$ and 
required $|m_{H_{u,d}}^2/m_{3/2}^2| < 5 $ which are determined 
by the electroweak symmetry breaking conditions in Eqs.\,(\ref{eq:tuning}) and (\ref{eq:angle}).
The ratios of the areas of each histogram roughly represent the relative 
consistency of the value of $\tan\beta$ in the pure gravity mediation.
}
\label{fig:L}
\end{center}
\end{figure}

By putting the typical values of $\tan\beta = O(1)$ 
and the Higgs mass parameters of the gravitino mass scale together
into the definition of $L$ in Eq.\,(\ref{eq:L}),  we find 
that the typical value of $L$ is also of the gravitino mass scale.
To see this clearly,
we show the typical size of $L$ for $\tan \beta = 1, 3, 10$ and $30$
(Fig.\,\ref{fig:L}).
Here, we have assumed that $\mu_H$ and $B$ range
from $m_{3/2}/3$ to $3m_{3/2}$, respectively.%
\footnote{
More precisely, we assumed that $\log_{10}\mu_H/m_{3/2}$ and $\log_{10}B/m_{3/2}$
obey the normal distribution with the mean value $0$ and the standard deviation 
$0.5\times\log_{10} 3$.
For a given $\tan\beta$, the Higgs squared masses are determined by  $m_{H_{u,(d)}}^2 = - |\mu_H^2| + B\mu_H \cot\beta^{(-1)} $.
In the figure, we generated the fixed number of random numbers for each $\tan\beta$.
Afterward, we required $|m_{H_{u,d}}^2/m_{3/2}^2|< 5$ so that they are  of 
the order of the gravitino mass.
Thus, the ratios of the areas of each histogram roughly represent the relative 
consistency of the value of $\tan\beta$ in the pure gravity mediation.
The figure shows that the model with $\tan\beta = O(10)$ is less consistent as expected.}
The figure also shows that $|L/m_{3/2}| \simeq 0.5-2$ for $\tan\beta = O(1)$.
Therefore, in the pure gravity mediation model, we expect $L/m_{3/2} = O(1)$, 
which leads to comparable contribution to the wino mass 
from the Higgsino threshold effects (see Eq.\,(\ref{eq:M2})).

\begin{figure}[t]
\begin{center}
\begin{minipage}{.49\linewidth}
  \includegraphics[width=.8\linewidth]{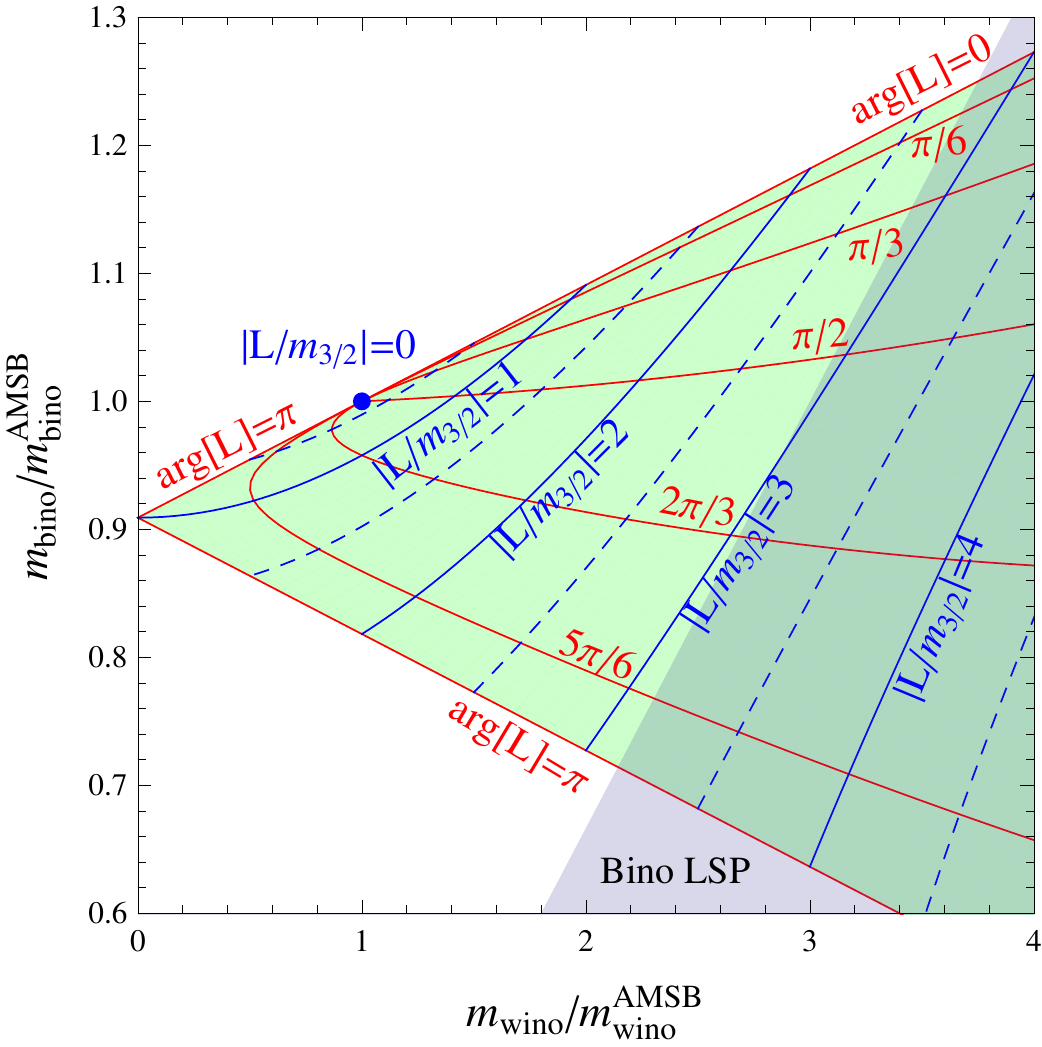}
\end{minipage}
\begin{minipage}{.49\linewidth}
  \includegraphics[width=.8\linewidth]{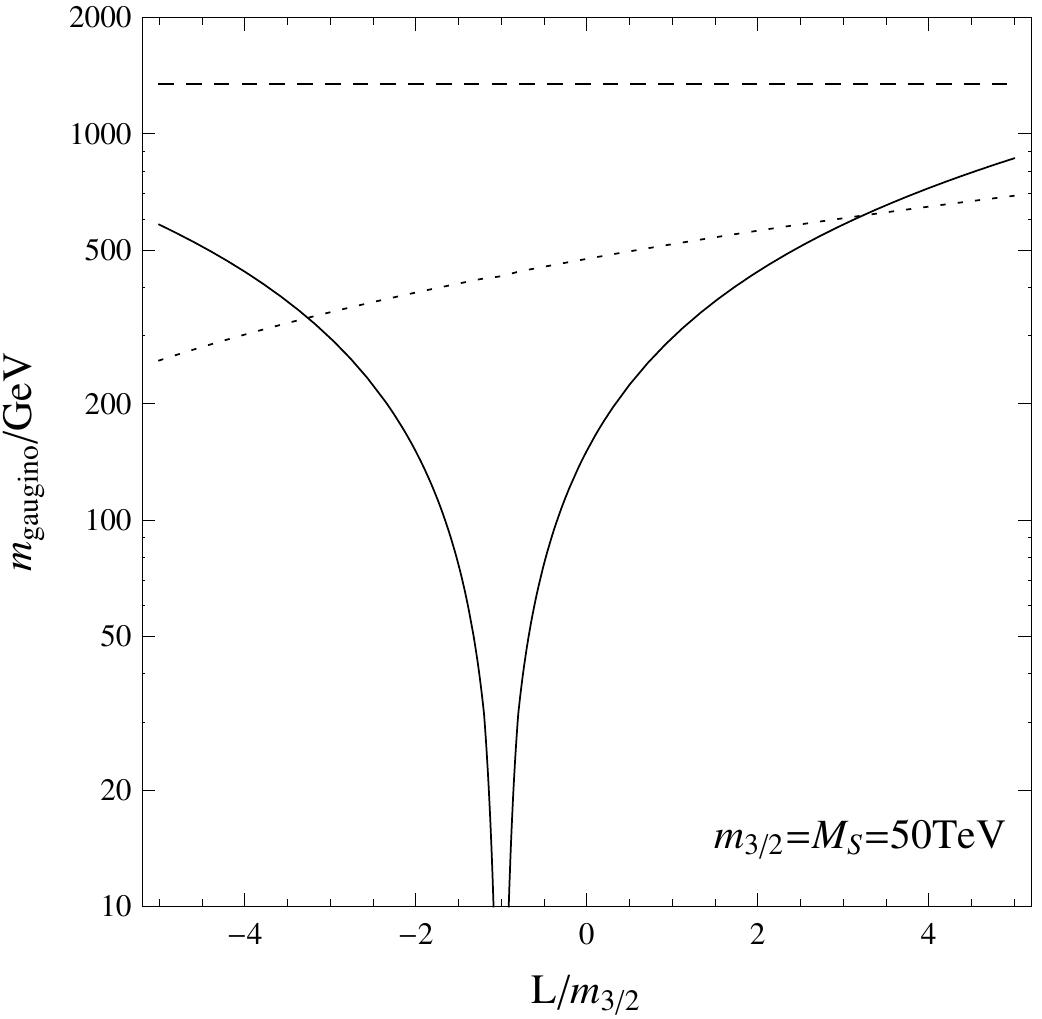}
  \end{minipage}
\caption{\sl \small
(Left) The ratios of the wino and bino masses with and without 
the Higgsino contributions for given values of $L$.
We have used a phase convention that $m_{3/2}$ is real and positive.
The red lines show the $|L|$ dependences for given phases of $L$, 
while the blue lines show the $\arg[L]$ dependences for given values of $|L|$.
(The dashed blue lines show the values of $|L|$ in between the ones for the two solid lines.).
In the gray shaded region for $|L/m_{3/2}|\gtrsim 3$, the wino is no more the LSP.
(Right)  The $L$ dependences of the gaugino masses for $m_{3/2}=M_{\rm SUSY}= 50$\,TeV
for $L>0\,(\arg[L] = 0)$ and $L<0\,(\arg[L] = \pi)$.
}
\label{fig:L-dependence}
\end{center}
\end{figure}

In Fig.\,\ref{fig:L-dependence}, we show the ratio of the wino and bino masses 
with and without the Higgsino contributions for given values of $L$ (left panel).
The figure shows that the wino mass can be about twice as heavy as the anomaly mediated 
contribution for $|L/m_{3/2}|\simeq 1$ which is expected in the pure gravity mediation model.
It should be noted that the wino becomes no more the LSP where the Higgsino 
threshold contribution dominates.
In such cases, the relic density of dark matter easily exceed the 
observed one due to the highly suppressed annihilation cross section of the bino for $O(100)\,$GeV.
Fortunately, however, the figure shows  that the bino becomes LSP only for $|L/m_{3/2}|>3$
which is less likely in the pure gravity mediation model.
Therefore, in the pure gravity mediation model,
 the LSP is mostly wino-like, although the wino mass obtains a comparable contribution
 from the Higgsino threshold effects.%
 \footnote{
In general, a relative phase between $L$ and $m_{3/2}$
is a free parameter, and hence, the three gauginos have different phases.
Such gaugino phases, however, do not cause serious CP-problems,
since the Higgsinos as well as the sfermions are expected to be very heavy
in the pure gravity mediation model. 
Interestingly, the relative phase of $O(1)$ may lead to
the visible electron electric dipole moment of $d_e/e \sim 10^{-30}$\,cm\,\cite{hep-ph/0409232}
for the $\mu$-term in the tens to hundreds TeV range,
which can be reached in future experiments\,\cite{EDM}.
}

\begin{figure}[t]
\begin{center}
  \includegraphics[width=.4\linewidth]{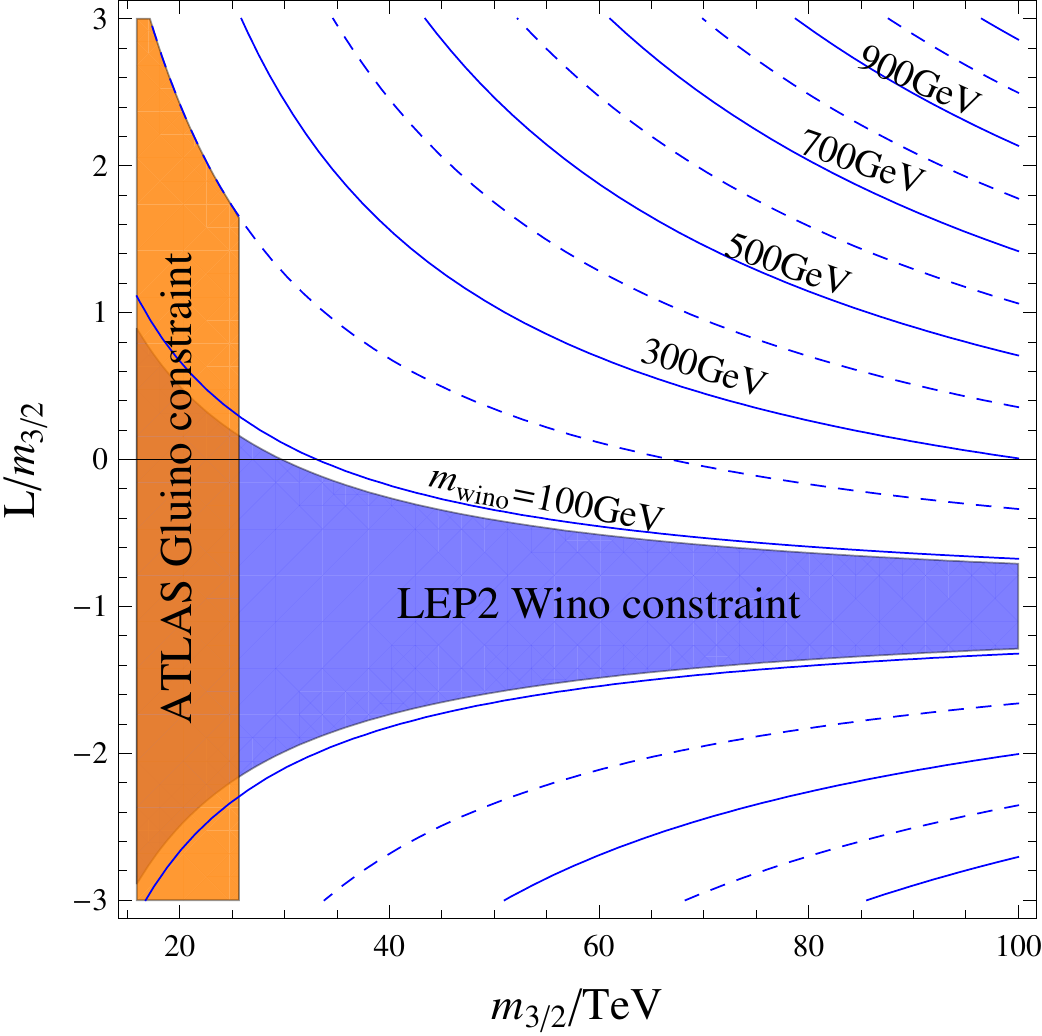}
\caption{\sl \small
The contour plot of the wino mass for $L>0$ and $L<0$.
Here, we have taken $M_{\rm SUSY} = m_{3/2}$ (blue lines).
(The dashed lines corresponds to $m_{\rm wino} = 150\,{\rm GeV},250\,{\rm GeV},\cdots$.)
The blue shaded region denotes the experimental
constraint,  $m_{\rm wino}\geq 88$\,GeV for the degenerated neutralino-chargino obtained 
by LEPII experiment\,\cite{hep-ex/0203020}.
The orange shaded region denotes the experimental
constrain on the gauginos, $m_{\rm gluino}\gtrsim 750$\,GeV for $m_{\rm LSP}<200$\,GeV
reported by the ATLAS collaboration\,\cite{ATLAS2}.
}
\label{fig:wino}
\end{center}
\end{figure}

\begin{figure}[t]
\begin{center}
  \includegraphics[width=.4\linewidth]{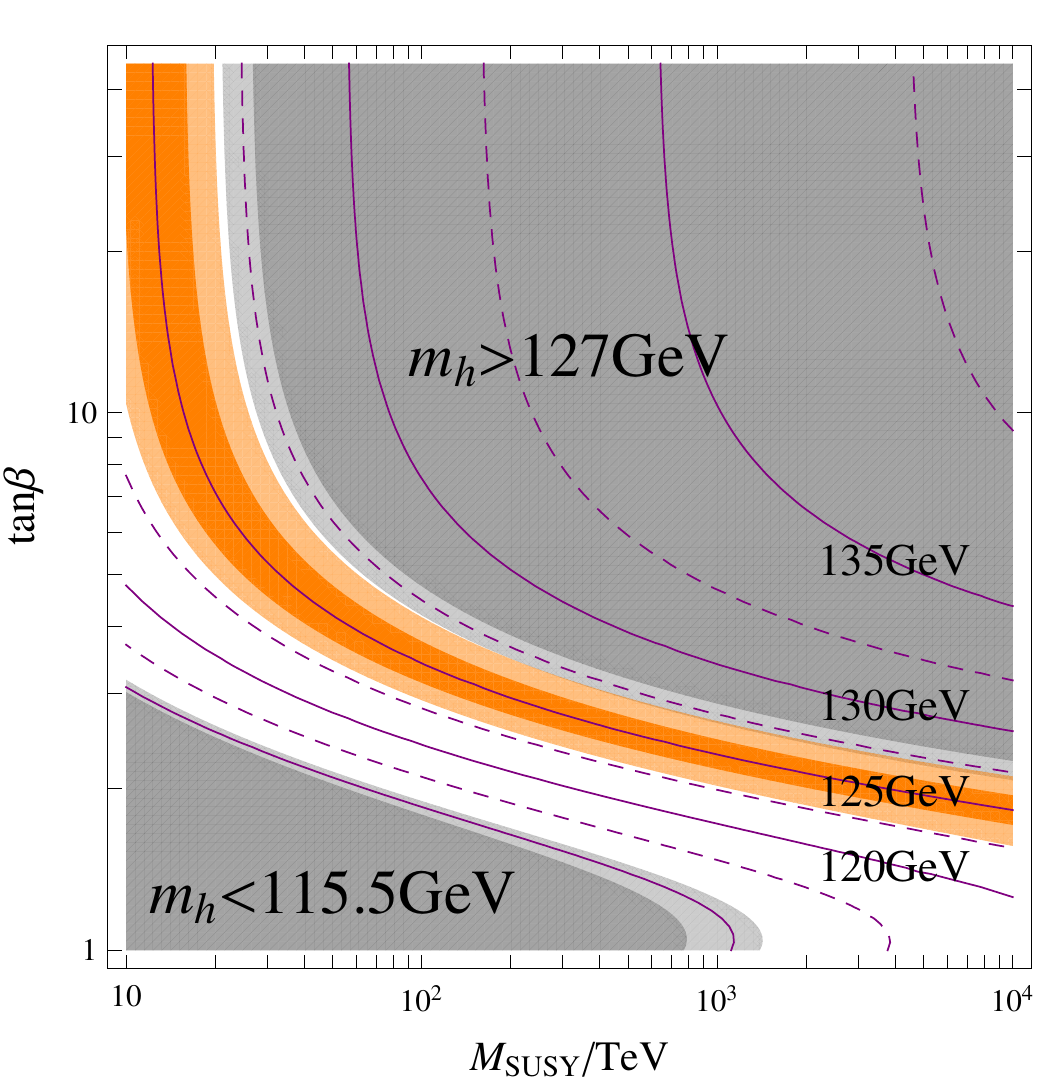}
\caption{\sl \small
The contour plot of the lightest Higgs boson mass.
(The dashed contours are for the intermediate 
values between the two solid contours.)
Here, we have fixed $m_{3/2}=50$\,TeV and taken $\mu_H=M_{\rm SUSY}$.
The gray shaded regions correspond to $m_h<115.5$\,GeV and $m_h >127$\,GeV 
which are excluded by the ATLAS and CMS collaborations
at 95\,\%\,C.L. for the central value of the top quark mass, $m_{\rm top} = 173.2\pm 0.9$\,GeV.
The light gray shaded region denotes the Higgs mass constraints including the $1\sigma$ error 
of the top quark mass.
The orange band shows the Higgs boson mass $124\,GeV < m_{h}< 126\,GeV$ hinted 
by the ATLAS and CMS collaborations for the central value of the top quark mass.
The light orange band is the one including the $1\sigma$ error of the top quark mass. 
}
\label{fig:Higgs}
\end{center}
\end{figure}

In Fig.\,\ref{fig:wino}, we show the contour plot of  the wino mass. 
In the figure, the blue shaded region 
shows the current experimental constraints 
on the wino mass   $m_{\rm wino}\geq 88$\,GeV for the degenerated neutralino-chargino obtained  gz
by LEPII experiment\,\cite{hep-ex/0203020}.
The orange shaded region shows the experimental constrain on the gauginos, 
$m_{\rm gluino}\gtrsim 750$\,GeV for $m_{\rm LSP}\lesssim 200$\,GeV reported by the ATLAS collaboration\,\cite{ATLAS2}. 
By remembering that $L/m_{3/2}\gtrsim 2.5$ is less likely in the pure gravity mediation, 
the figure shows that the gluino mass bound requires $m_{3/2}\gtrsim 30$\,TeV.%
\footnote{
Fig.\,\ref{fig:L} shows that $L/m_{3/2}\gtrsim 2.5$ is possible for $\tan\beta\simeq 1$. 
As we will see from Fig.\,\ref{fig:Higgs}, however, 
the lightest Higgs boson mass of our main concern ($124$\,GeV$<m_h<126$\,GeV) 
requires $m_{3/2} \gg 100$\,TeV for $\tan\beta \simeq 1$.
Thus, the conclusion $m_{3/2}\gtrsim 30$\,GeV is not changed. 
}

Finally, we discuss the lightest Higgs boson mass in the pure gravity mediation model.
In the pure gravity mediation model,
the lightest Higgs boson mass is expected to be heavier than
the conventional MSSM models due to 
the heavy scalar bosons\,\cite{TU-363}.
In Fig.\,\ref{fig:Higgs},  we show the Higgs boson mass obtained 
by solving the full one-loop renormalization-group
equations of the Higgs quartic coupling
and other coupling constants given in Ref.\,\cite{hep-ph/0406088}
with the boundary condition, 
\begin{eqnarray}
\label{eq:SUSY}
  \lambda = \frac{1}{4} \left(\frac{3}{5}g_1^2+ g_2^2 \right) \cos^22\beta\ ,
\end{eqnarray}
at the heavy scalar scale.
The threshold corrections at the heavy scalar scale are also taken into account.
We also take into account the weak scale threshold corrections to those 
parameters in accordance with Ref.\,\cite{arXiv:0705.1496,arXiv:1108.6077}.
It should be noted that the predicted Higgs boson mass is slightly lighter than 
the one in Ref.\,\cite{arXiv:1108.6077} for a given $(M_{\rm SUSY},\tan\beta)$,
since the Higgsino contributions decouple at the very high scale in the pure gravitino mediation model
(see Ref.\,\cite{Ibe:2011aa}).

In the figure, the gray shaded regions correspond to $m_h<115.5$\,GeV and $m_h >127$\,GeV 
which are excluded by the ATLAS and CMS collaborations\,\cite{ATLAS,CMS}
at 95\,\%\,C.L. for the central value of the top quark mass, 
$m_{\rm top} = 173.2\pm 0.9$\,GeV\,\cite{arXiv:1107.5255}.
The light gray shaded region denotes the Higgs mass constraints including the $1\sigma$ error 
of the top quark mass.
The orange band shows the Higgs boson mass $124\,{\rm GeV}<m_{h}< 126\,{\rm GeV}$ hinted 
by the ATLAS and CMS collaborations\,\cite{ATLAS,CMS} for the central value of the top quark mass.
The light orange band is the one including the $1\sigma$ error of the top quark mass.

By combined with $m_{3/2}\gtrsim 30$\,TeV which is required from 
the experimental gluino mass bound,
the hinted Higgs boson in the Fig.\,\ref{fig:Higgs} 
($124\,{\rm GeV}<m_{h}<126\,{\rm GeV}$) constrains the value of $\tan\beta$ to $\tan\beta \lesssim 7$.
This shows that the pure gravity mediation works quite consistently since 
$\tan\beta = O(1)$ is expected in the pure gravity mediation model.

\section{Signals of the pure gravity mediation model}\label{sec:signals}
In this section, we consider several signals predicted in the pure gravity mediation model. Before going to discuss those, we summarize current cosmological constraints on the model. After that, we consider signals related to dark matter detections, where current astrophysical constraints on the dark matter mass and near-future prospects to detect the dark matter are discussed. We finally consider collider signals with particularly focusing on the pair production of the gluino at the LHC 
experiments with the center of mass energy of 14\,TeV.

\subsection{Cosmological constraints}
We first consider the thermal history of the dark matter which is the neutral wino in the pure gravity mediation model. Its $SU(2)_L$ partner, the charged wino, is slightly heavier than the neutral one by $155$--$170\,$MeV because of contributions from one-loop gauge boson diagrams\,\cite{Cheng:1998hc}. The charged wino decays into a neutral wino and a pion with the lifetime of ${\cal O}(10^{-10})$\,sec. 
It is known that the thermal relic density of the wino, which is obtained by considering not only self-annihilation processes of the neutral wino but also co-annihilation processes between the 
neutral and/or the charged winos, can be consistent with the observed dark matter density when its mass is $m_{\rm wino} \simeq 2.7$\,TeV. This is because the annihilation cross section of the wino is highly boosted by the non-perturbative effect called Sommerfeld-enhancement\,\cite{Hisano:2006nn}.

On the other hand, the wino dark matter is also produced non-thermally through the late time decay of the gravitino, which also contributes to the relic abundance of the dark matter. If the contribution is significant, the neutral wino consistent with the observed dark matter density is much lighter than 
$2.7\,$TeV\,\cite{Gherghetta:1999sw,hep-ph/9906527}. 
 In particular, in order to have an appropriate reheating temperature for the successful thermal leptogenesis, there is an upper bound on the wino mass; 
 $m_{\rm wino} \lesssim 1$\,TeV\,\cite{Ibe:2011aa}. 
 This fact means that the most of the dark matter observed today is not from thermal relics but produced non-thermally by the late time decay of the gravitino.

Since the neutral wino has a large annihilation cross section into a $W$-boson pair, which is of the order of $10^{-24}$--$10^{-25}$ cm$^3$/s when $m_{\rm wino} \lesssim 1$\,TeV, it may affect several phenomena in the early universe~\cite{Moroi:2011ab}. For instance, the annihilation may affect abundances of light elements, and, in fact, observations of the elements put a bound on the mass of the neutral wino as $m_{\rm wino} \gtrsim 200$\,GeV in order not to destroy the elements during Big-Bang Nucleosynthesis (BBN)\,\cite{DM_BBN}. The annihilation also affects the recombination history of the universe. If the annihilation is significantly large, it modifies the spectrum of cosmic microwave background\,\cite{DM_CMB}. This fact leads to the constraint as $m_{\rm wino} \gtrsim 200$\,GeV, 
which is comparable to that from BBN.

\subsection{Dark matter detections}
Since the $\mu$-parameter is of the order of 10--100\,TeV in the pure gravity mediation model, the effect of the mixing between wino and higgsino components on the lightest supersymmetric particle (dark matter) is negligibly small. The scattering cross section between the dark matter and a nucleon is then estimated to be $10^{-47}$ cm$^2$\,\cite{Hisano:2010fy}, which seems to be very challenging to discover the dark matter in on-going direct detection experiments. This is a sharp contrast to the cases of Split Supersymmetry model and conventional anomaly mediation models. Since the $\mu$-parameter does not have to be huge in these models, the tree-level diagram that the higgs boson is exchanged in the $t$-channel contributes to the scattering cross section significantly, which enables us to detect the dark matter in near future\,\cite{Moroi:2011ab}. Direct detection experiments of dark matter can be therefore used as a test of the pure gravity mediation model.

\begin{figure}[t]
\begin{center}
\includegraphics[width=0.5\linewidth]{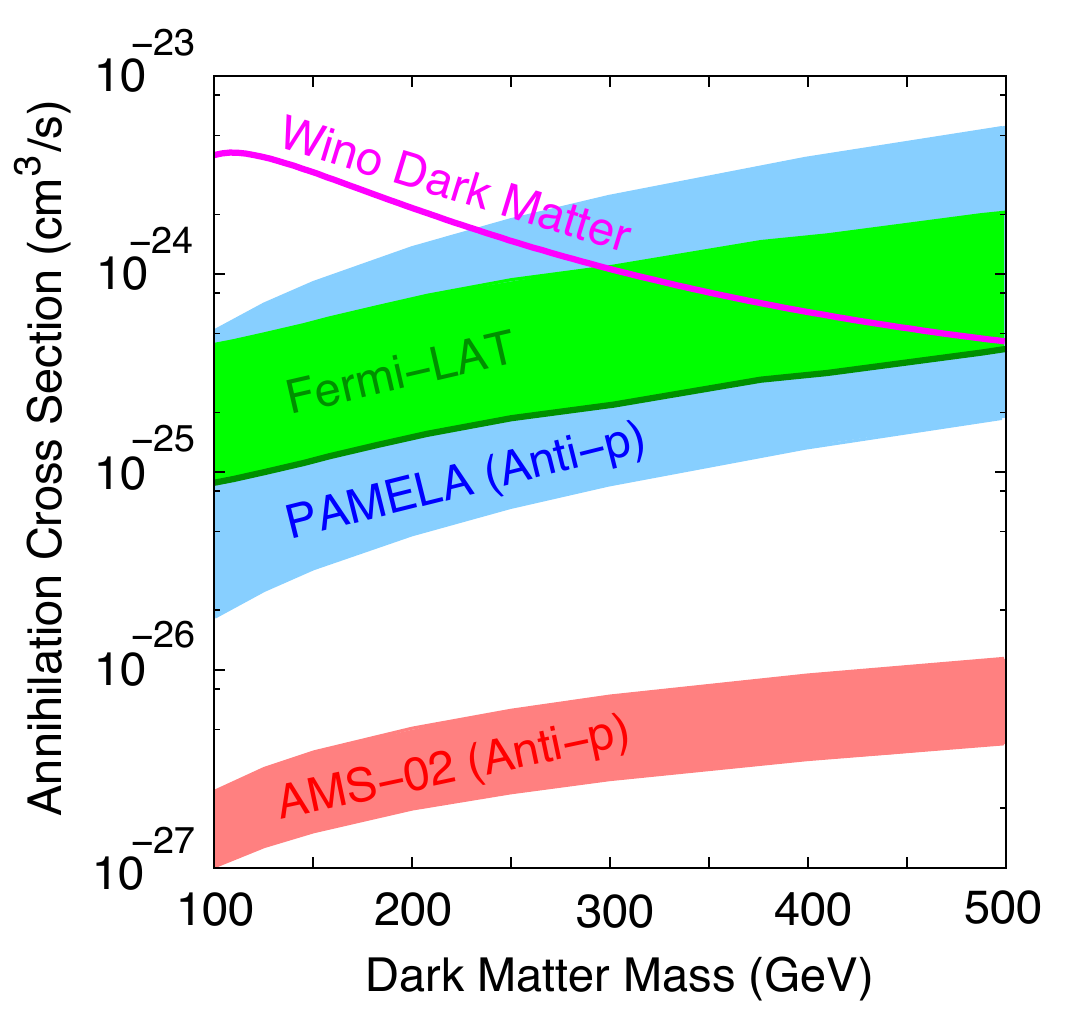}
\caption{\sl\small Constraints and future prospects of indirect detection experiments of dark matter. Theoretical prediction of the neutral wino dark matter is also shown.}
\label{fig: DM}
\end{center}
\end{figure}

On the contrary to the direct detection of dark matter, we can expect rich signals at indirect detection experiments, because the dark matter is almost purely wino in the pure gravity mediation model and its annihilation cross section is boosted by the Sommerfeld effect\,\cite{Sommerfeld}. Among several on-going experiments, the most stringent constraint on the dark matter is obtained by the Fermi-LAT experiment observing gamma-rays from milky way satellites\,\cite{Ackermann:2011wa}. This constraint is depicted in Fig.\,\ref{fig: DM} as a solid (green) line. No astrophysical boost factor is assumed here. Theoretical prediction of the neutral wino is also shown in the figure, which is obtained by calculating its annihilation cross section involving the Sommerfeld effect at one-loop level\,\cite{Hryczuk:2011vi}. 
Notice, however, that there may be some uncertainties on the constraint, 
since the constraint is based on several assumptions such as the use of fixed dark matter profile. 
According to Ref.\,\cite{Charbonnier:2011ft} in which those uncertainties (involving dark matter profiles) on the gamma-ray experiment are discussed, we also show the region (green-shaded one) above the constraint in order to take the uncertainties into account. It can be seen that the neutral wino should be, at least, heavier than $300$\,GeV.

Another interesting indirect detection is the PAMELA experiment observing the cosmic-ray $\bar{p}$ (anti-proton) flux\,\cite{Adriani:2010rc}. Current constraint on the dark matter from the experiment is also shown in Fig.\,\ref{fig: DM} as a blue-shaded region. Since the $\bar{p}$ flux depends on how $\bar{p}$ propagates under the complicated magnetic field of our galaxy and which dark matter profiles we 
adopt\,\cite{Evoli:2011id}, the constraint has large uncertainties as can be seen in the figure. The mass of the dark matter is, however, constrained to be $m_{\rm wino} \gtrsim 230$\,GeV in spite of the uncertainties. On the other hand, the observation of the cosmic-ray $\bar{p}$ flux in near future is very hopeful. This is because the AMS-02 experiment, which has already been started\,\cite{AMS-02}, has better sensitivity than the PAMELA experiment and it is also expected that astrophysical uncertainties related to the $\bar{p}$ propagation are reduced. The future sensitivity to detect the dark matter in this experiment is also depicted in the figure as a red-shaded region with assuming an appropriate propagation model\,\cite{Evoli:2011id}. It can be seen that the sensitivity is much below the prediction of the dark matter. It is also worth noting that the whole mass range of the dark matter consistent with the thermal leptogenesis will be fully tested by the future observation of the cosmic-ray $\bar{p}$ flux, because the annihilation cross section of the dark matter is not suppressed because of the Sommerfeld effect. It may be even possible to determine $m_{\rm wino}$ by observing the $\bar{p}$ spectrum.

Finally, we comment on other indirect detections of dark matter. It is well known that there is an anomaly at the cosmic-ray $e^+$ flux\,\cite{Adriani:2008zr}. Since it is difficult to account for the anomaly by the neutral wino dark matter with the mass of 300--1000 GeV~\cite{PAMELA WINO}, it should be explained by some astrophysical activities. The observation of the $e^+$ flux is therefore not better than that of the $\bar{p}$ flux to test the pure gravity mediation model. The observation of the $\nu$ flux from the galactic center may give an good opportunities to test the neutral wino dark matter\,\cite{Moroi:2011ab}, though the signal strength depends on the dark matter profile at the center. On the other hand, the observation of the $\nu$ flux from the sun seems to be challenging, because the flux is proportional to the spin-dependent scattering cross section of the dark matter and it is estimated to be as small as 
$10^{-48}$\,cm$^2$ in the pure gravity mediation model\,\cite{Hisano:2010fy}.

\subsection{Collider signals}

\begin{figure}[t]
\begin{center}
\includegraphics[width=0.45\linewidth]{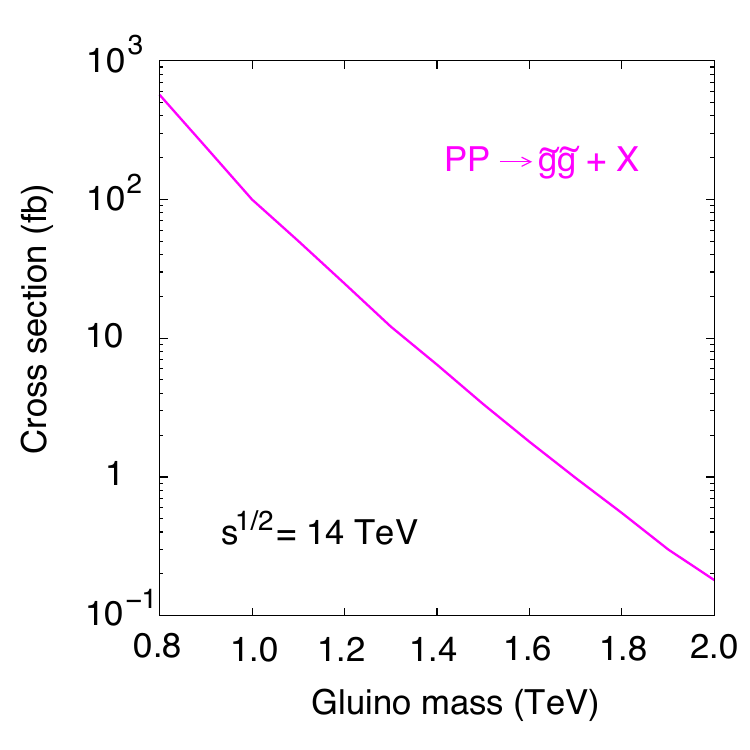}
\qquad
\includegraphics[width=0.435\linewidth]{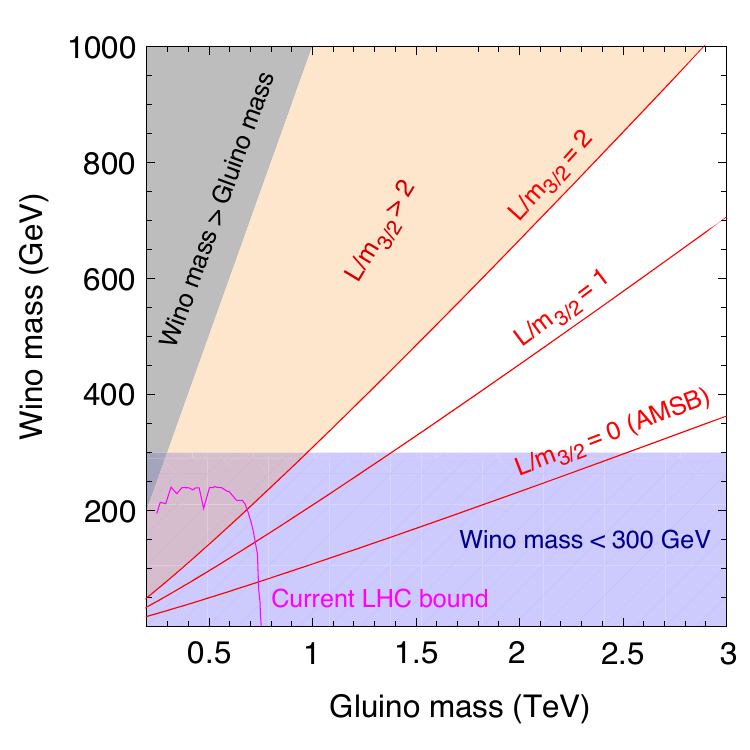}
\caption{\sl\small (Left panel) Cross section of the gluino pair production at the LHC experiment with the center of mass energy of 14\,TeV. (Right panel) Gluino and wino masses within the parameter region of $m_{\rm gluino} \lesssim 3$\,TeV. Shaded regions are not favored because of the gluino LSP ($m_{\rm gluino} > m_{\rm wino}$), too large $L$ ($L/m_{3/2} > 2$), and dark matter constraints ($m_{\rm wino} < 300$\,GeV). Current bound of the LHC experiment (7\,TeV) is also shown.}
\label{fig: LHC}
\end{center}
\end{figure}

In the pure gravity mediation model, the ratio between gluino and wino masses can be smaller than that of the conventional anomaly mediation model. The gluino may therefore be produced at the LHC experiment even if the wino mass is constrained to be $m_{\rm wino} \gtrsim 300$\,GeV. On the other hand, all the sfermions as well as the higgsinos are of the order of $10$--$100$\,TeV in the model and they are never produced at the LHC. 
As a result, the dominant collider signal of the model is  the pair production of the gluinos, whose production cross section is shown in Fig.\,\ref{fig: LHC} (left panel). Once the gluino is produced, it eventually decays into a neutral wino by emitting Standard Model particles. 
It is known that, when the sfermions are much heavier than the gluino, 
the radiative decay of the gluino into a gluon and a neutralino ($\tilde{g} \to g\tilde{\chi}^0$) 
can have a sizable branching fraction\,\cite{Radiative gluino decay}. In the pure gravity mediation model, however, the $\mu$-parameter is also as large as the sfermion masses and the branching fraction is much suppressed. The non-observation of the radiative decay therefore enables us to distinguish the pure gravity mediation model from other models predicting heavy sfermions without the large $\mu$-parameter.

Gluinos in the pure gravity mediation model therefore decay into two quarks with a neutralino/chargino ($\tilde{g} \to q\bar{q}^\prime + \tilde{\chi}^0/\tilde{\chi}^\pm_i$). The chargino, which is nothing but the charged wino, decays into a neutral wino (dark matter) by emitting a soft pion. On the other hand, when the neutralino is the bino, it decays through several modes; a charged wino $+$ a $W$-boson ($\tilde{B} \to \tilde{W}^\pm W^\mp$), a neutral wino $+$ a higgs boson ($\tilde{B} \to \tilde{W}^0 h$), or a charged/neutral wino $+$ two leptons ($\tilde{B} \rightarrow \tilde{W} l \bar{l}^\prime$), whose branching fractions depend highly on model parameters. In Fig.\,\ref{fig: LHC} (right panel), we show the range of gluino and wino masses within the parameter region of our interest for the LHC experiment. It is also worth noting that, as shown in the previous section, the bino mass is roughly given by $m_{\rm bino} \simeq m_{\rm gluino}/3$ in the most parameter region. 
Thus, the mass degeneracy between gluino and neutralino/chargino is not severe, which is very attractive from the viewpoint of discovering the signal.

The most efficient mode to discover the signal of the pure gravity mediation model is the pair production of the gluinos followed by the decay $\tilde{g} \to q\bar{q}^\prime \tilde{\chi}$ with $q (q^\prime)$ being a quark except the top quark and $\tilde{\chi}$ being a neutralino/chargino, so that the signal event is composed of four jets + missing energy. The branching fraction of the decay is about 73.4\,\% when all squark masses are degenerated. In Fig.\,\ref{fig: LHC} (right panel), the current bound on the $(m_{\rm gluino}, m_{\rm wino})$-plane, which is obtained by the LHC experiment with the center of mass energy of $7$\,TeV and 1.04\,fb$^{-1}$ data, is depicted with assuming that $m_{\rm wino} \sim m_{\rm bino}$ and 100\,\% branching fraction of the decay $\tilde{g} \to q\bar{q}^\prime \tilde{\chi}$\,\cite{ATLAS-CONF}. It can be seen that the region constrained by the current LHC data has already been excluded by dark matter experiments. It has been also shown that the gluino mass up to 1.2\,TeV can be discovered at the LHC experiment with the center of mass energy of 14\,TeV when 10\,fb$^{-1}$ data is accumulated\,\cite{Asai:2007sw}.

Once the signal of the pure gravity mediation model is discovered, the next important task will be the mass determinations of gauginos. When the gluino is about 1\,TeV and 100\,fb$^{-1}$ data is accumulated at the LHC experiment with the center of mass energy of 14\,TeV, the mass difference between gluino and wino can be determined with the accuracy of 5\,\%, which is obtained by observing the endpoint of two-jets invariant mass distribution at "four jets + missing energy" events\,\cite{Asai:2007sw}. The mass difference will be determined more accurately with the use of a novel method recently proposed in Ref.\,\cite{ISR}, where the endpoint of so-called $M_{T2}$ distribution\,\cite{Barr:2003rg} is shown to be stable against the contamination of initial state radiations. On the other hand, the gluino mass may be determined by observing the cross section of the gluino pair production if the acceptance of the LHC experiment for this mode is well understood. The wino mass is expected to be determined by observing the $M_{T2}$ endpoint, because the endpoint has a kink structure at the wino mass as a function of the test mass defining $M_{T2}$\,\cite{ISR}. It has been also shown that the wino mass is determined by using the charged track of $\tilde{W}^{\pm}$, because its decay length is estimated to be ${\cal O}(10)$\,cm\,\cite{Asai:2008sk}. 
It may be even possible to measure the lifetime of $\tilde{W}^{\pm}$ using this method. 
The mass difference between bino and wino is determined only when the branching fraction ${\rm Br}(\tilde{g} \to q\bar{q}\tilde{B}) \times {\rm Br}(\tilde{B} \rightarrow l\bar{l}\tilde{W}^0)$ is large 
enough\,\cite{Asai:2007sw}.

Finally, we comment on collider signals of the pure gauge mediation model when the gluino is heavier than a few TeV and is not accessible at the current and near future LHC experiments. In such cases,
we have to rely on the direct production (Drell-Yan process) of charged winos associated with a quark (gluon). Associated quark (gluon) is necessary as a trigger for recording data\,\cite{Ibe:2006de, Moroi:2011ab}. Its cross section is rapidly decreased with increasing the wino mass. 
Since the mass of the wino is at most 1\,TeV for the successful leptogenesis, the high luminosity LHC experiment (HL-LHC)\,\cite{HL-LHC} may help us to discover the signal. 
On the other hand, if the multi-TeV linear colliders such as ILC~\cite{ILC} or CLIC~\cite{CLIC} are available, we can investigate the properties of neutral and charged wino in details. 
Since the analysis strategy for the mode $e^+e^- \to \tilde{W}^+\tilde{W}^- \to \tilde{W}^0\tilde{W}^0\pi^+\pi^-$ is very similar to that for the golden mode of dark matter detections, $e^+e^- \to \tilde{\chi}^+\tilde{\chi}^- \to \tilde{\chi}^0\tilde{\chi}^0W^+W^-$ with $\chi^0$ and $\chi^\pm$ being the dark matter and its charged partner\,\cite{Asano:2011aj}, 
we can easily find the signal of the pure gravity mediation model if $\pi$-mesons are efficiently detected.

\section{Conclusion}\label{sec:conclusion}
The pure gravity mediation model is the bare bones model of the supersymmetric
Standard Model.
Despite its simpleness, the model is quite successful for $m_{3/2} = O(10-10^2)$\,TeV;
the model has a good candidate of dark matter, 
the gauge coupling constants unify at the GUT scale very precisely, 
the Higgs boson mass around $125$\,GeV can be easily accounted.
In this sense, the model is superior even to the Standard Model.
The consistency with the thermal leptogenesis is also an significant 
support of the model.

In this paper, we discussed details of the gaugino mass spectrum in the pure 
gravity mediation model.
There, we showed that the wino mass obtains comparable contributions 
both from the anomaly-mediation and the Higgsino threshold effects.
As a result, the ratio between the wino LSP and the gluino masses 
can be as large as around one third, which enhances the detectability
of the model at the LHC experiments.
In fact, we showed that the gluino can be within the reach of the LHC experiments
even for the wino mass which satisfies the cosmological and astrophysical constraints, 
$m_{\rm wino}\gtrsim 300$\,GeV.
This is a sharp contrast to the cases of the anomaly mediated gaugino spectrum 
where the gluino mass is about eight to nigh times larger than the wino mass.
Utilizing this property, we discussed the strategies of the discovery and the measurement 
of the model at the LHC experiments via the gluino production.

In this paper, we also discussed the prospects of the wino dark matter detection 
via  the cosmic ray observations.
As a result, we found that the wino dark matter scenario 
which is consistent with the thermal leptogenesis 
can be fully surveyed by observing the cosmic ray  anti-proton flux
at the AMS-02 experiment.
Therefore, the most motivated parameter region of the pure gravity mediation 
model which is consistent with the thermal leptogenesis
can be tested over the next ten years or so.

\section*{Acknowledgments}
This work is supported by Grant-in-Aid for Scientific research from the Ministry of Education, Science, Sports, and Culture (MEXT), Japan, No.\ 22244021 (S.M. and T.T.Y.) and No.\ 23740169 (S.M.), and also by World Premier International Research Center Initiative (WPI Initiative), MEXT, Japan.


\begin{thebibliography}{99}

\bibitem{Ibe:2011aa} 
  M.~Ibe and T.~T.~Yanagida,
  arXiv:1112.2462 [hep-ph].


\bibitem{Nilles:1983ge}
For  a review, H.~P.~Nilles,
  Phys.\ Rept.\  {\bf 110} (1984) 1.
  
\bibitem{Inoue:1991rk}
  K.~Inoue, M.~Kawasaki, M.~Yamaguchi and T.~Yanagida,
  Phys.\ Rev.\ D {\bf 45}, 328 (1992).

  
\bibitem{Giudice:1998xp}
  G.~F.~Giudice, M.~A.~Luty, H.~Murayama and R.~Rattazzi,
  %
  JHEP {\bf 9812}, 027 (1998).
\bibitem{Randall:1998uk}
  L.~Randall and R.~Sundrum,
  %
  Nucl.\ Phys.\ B {\bf 557}, 79 (1999).
\bibitem{hep-ph/9205227} 
  M.~Dine and D.~MacIntire,
  Phys.\ Rev.\ D\ {\bf 46}, 2594  (1992)
  [hep-ph/9205227].


\bibitem{Pagels:1981ke} 
  H.~Pagels and J.~R.~Primack,
  Phys.\ Rev.\ Lett.\  {\bf 48}, 223 (1982);
  S.~Weinberg,
  Phys.\ Rev.\ Lett.\  {\bf 48}, 1303 (1982);
  M.~Y.~.Khlopov and A.~D.~Linde,
  Phys.\ Lett.\ B {\bf 138}, 265 (1984).
  
\bibitem{kkm}
 M.~Kawasaki, K.~Kohri and T.~Moroi,
 Phys.\ Rev.\ D {\bf 71} (2005) 083502
 [arXiv:astro-ph/0408426]; 
  K.~Jedamzik,
  Phys.\ Rev.\  D {\bf 74}, 103509 (2006)
  [arXiv:hep-ph/0604251];
  M.~Kawasaki, K.~Kohri, T.~Moroi and A.~Yotsuyanagi,
  Phys.\ Rev.\ D {\bf 78}, 065011 (2008)
  [arXiv:0804.3745 [hep-ph]],
 and references therein.
 
\bibitem{leptogenesis}
  M.~Fukugita and T.~Yanagida,
  Phys.~Lett.~{\bf B174} (1986) 45; 
  For  reviews,
  W.~Buchmuller, P.~Di Bari and M.~Plumacher,
  Annals Phys.\  {\bf 315}, 305 (2005)
  [hep-ph/0401240];
  W.~Buchmuller, R.~D.~Peccei and T.~Yanagida,
  Ann.\ Rev.\ Nucl.\ Part.\ Sci.\  {\bf 55}, 311 (2005)
  [arXiv:hep-ph/0502169];  
  S.~Davidson, E.~Nardi and Y.~Nir,
  Phys.\ Rept.\ \ {\bf 466}, 105  (2008)
  [arXiv:0802.2962 [hep-ph]].


\bibitem{Gherghetta:1999sw}
  T.~Gherghetta, G.~F.~Giudice and J.~D.~Wells,
  Nucl.\ Phys.\  B {\bf 559}, 27 (1999)
  [arXiv:hep-ph/9904378].
\bibitem{hep-ph/9906527} 
  T.~Moroi and L.~Randall,
  Nucl.\ Phys.\ B\ {\bf 570}, 455  (2000)
  [hep-ph/9906527].
 
\bibitem{Ibe:2004tg}
  M.~Ibe, R.~Kitano, H.~Murayama and T.~Yanagida,
  Phys.\ Rev.\  D {\bf 70}, 075012 (2004)
  [arXiv:hep-ph/0403198];
  M.~Ibe, R.~Kitano and H.~Murayama,
  Phys.\ Rev.\  D {\bf 71}, 075003 (2005)
  [arXiv:hep-ph/0412200].


\bibitem{Polonyi}
  G.~D.~Coughlan, W.~Fischler, E.~W.~Kolb, S.~Raby and G.~G.~Ross,
  Phys.\ Lett.\ B {\bf 131}, 59 (1983).

\bibitem{hep-ph/0605252} 
  M.~Ibe, Y.~Shinbara and T.~T.~Yanagida,
  Phys.\ Lett.\ B\ {\bf 639}, 534  (2006)
  [hep-ph/0605252].
\bibitem{Abe:2011ts} 
  K.~Abe, T.~Abe, H.~Aihara, Y.~Fukuda, Y.~Hayato, K.~Huang, A.~K.~Ichikawa and M.~Ikeda {\it et al.},
  arXiv:1109.3262 [hep-ex].
  
\bibitem{ATLAS}  
  ATLAS report, ATLAS-CONF-2011-163.
\bibitem{CMS}
  CMS~Collaboration,
  arXiv:1202.1487 [hep-ex].

\bibitem{Ibe:2006de}
M.~Ibe, T.~Moroi and T.~T.~Yanagida,
Phys.\ Lett.\  B {\bf 644}, 355 (2007)
[arXiv:hep-ph/0610277].
\bibitem{Giudice:1988yz} 
  G.~F.~Giudice and A.~Masiero,
  Phys.\ Lett.\ B {\bf 206}, 480 (1988).
  
\bibitem{hep-th/0405159} 
  N.~Arkani-Hamed and S.~Dimopoulos,
  JHEP\ {\bf 0506}, 073  (2005)
  [hep-th/0405159].
\bibitem{hep-ph/0406088} 
  G.~F.~Giudice and A.~Romanino,
  Nucl.\ Phys.\ B\ {\bf 699}, 65  (2004)
  [Erratum-ibid.\ B\ {\bf 706}, 65  (2005)]
  [hep-ph/0406088].
\bibitem{hep-ph/0409232} 
  N.~Arkani-Hamed, S.~Dimopoulos, G.~F.~Giudice and A.~Romanino,
  Nucl.\ Phys.\ B\ {\bf 709}, 3  (2005)
  [hep-ph/0409232].
\bibitem{Izawa:2010ym} 
  K.~-I.~Izawa, T.~Kugo and T.~T.~Yanagida,
  Prog.\ Theor.\ Phys.\  {\bf 125}, 261 (2011)
  [arXiv:1008.4641 [hep-ph]].


\bibitem{hep-ph/0411041} 
  J.~D.~Wells,
  Phys.\ Rev.\ D\ {\bf 71}, 015013  (2005)
  [hep-ph/0411041].
\bibitem{arXiv:1111.4519} 
  L.~J.~Hall and Y.~Nomura,
  arXiv:1111.4519 [hep-ph].
  \bibitem{EDM}
  See e.g. 
Amar C. Vutha, Wesley C. Campbell, Yulia V. Gurevich, Nicholas R. Hutzler, Maxwell Parsons, David Patterson, Elizabeth Petrik, Benjamin Spaun, John M. Doyle, Gerald Gabrielse, David DeMille,
arXiv:0908.2412\,[physics].
\bibitem{hep-ex/0203020} 
  A.~Heister {\it et al.} [ALEPH Collaboration],
  Phys.\ Lett.\ B\ {\bf 533}, 223  (2002)
  [hep-ex/0203020].

\bibitem{ATLAS2}  
ATLAS reprot,  ATLAS-CONF-2011-155
  
\bibitem{TU-363} 
  Y.~Okada, M.~Yamaguchi and T.~Yanagida,
  Phys.\ Lett.\ B\ {\bf 262}, 54  (1991).
  


\bibitem{arXiv:0705.1496} 
  N.~Bernal, A.~Djouadi and P.~Slavich,
  JHEP\ {\bf 0707}, 016  (2007)
  [arXiv:0705.1496 [hep-ph]].
  
\bibitem{arXiv:1108.6077} 
  G.~F.~Giudice and A.~Strumia,
  Nucl.\ Phys.\ B {\bf 858}, 63 (2012)
  [arXiv:1108.6077 [hep-ph]].
   \bibitem{arXiv:1107.5255} 
  M.~Lancaster [Tevatron Electroweak Working Group and for the CDF and D0 Collaborations],
  arXiv:1107.5255 [hep-ex].
 


\bibitem{Cheng:1998hc}
H.~C.~Cheng, B.~A.~Dobrescu and K.~T.~Matchev,
Nucl.\ Phys.\  B {\bf 543}, 47 (1999)
[arXiv:hep-ph/9811316].

\bibitem{Hisano:2006nn}
J.~Hisano, S.~Matsumoto, M.~Nagai, O.~Saito and M.~Senami,
Phys.\ Lett.\  B {\bf 646}, 34 (2007)
[arXiv:hep-ph/0610249].

\bibitem{Moroi:2011ab}
T.~Moroi and K.~Nakayama,
arXiv:1112.3123 [hep-ph].

\bibitem{DM_BBN}
K.~Jedamzik,
Phys.\ Rev.\  D {\bf 70}, 083510 (2004)
[arXiv:astro-ph/0405583];
J.~Hisano, M.~Kawasaki, K.~Kohri and K.~Nakayama,
Phys.\ Rev.\  D {\bf 79}, 063514 (2009)
[Erratum-ibid.\  D {\bf 80}, 029907 (2009)]
[arXiv:0810.1892 [hep-ph]];
J.~Hisano, M.~Kawasaki, K.~Kohri, T.~Moroi and K.~Nakayama,
Phys.\ Rev.\  D {\bf 79}, 083522 (2009)
[arXiv:0901.3582 [hep-ph]].

\bibitem{DM_CMB}
B.~Ezhuthachan, S.~Mukhi and C.~Papageorgakis,
JHEP {\bf 0904}, 101 (2009)
[arXiv:0903.0003 [hep-th]];
T.~R.~Slatyer, N.~Padmanabhan and D.~P.~Finkbeiner,
Phys.\ Rev.\  D {\bf 80}, 043526 (2009)
[arXiv:0906.1197 [astro-ph.CO]];
T.~Kanzaki, M.~Kawasaki and K.~Nakayama,
Prog.\ Theor.\ Phys.\  {\bf 123}, 853 (2010)
[arXiv:0907.3985 [astro-ph.CO]];
J.~Hisano, M.~Kawasaki, K.~Kohri, T.~Moroi, K.~Nakayama and T.~Sekiguchi,
Phys.\ Rev.\  D {\bf 83}, 123511 (2011)
[arXiv:1102.4658 [hep-ph]];
S.~Galli, F.~Iocco, G.~Bertone and A.~Melchiorri,
Phys.\ Rev.\  D {\bf 84}, 027302 (2011)
[arXiv:1106.1528 [astro-ph.CO]].

\bibitem{Hisano:2010fy}
J.~Hisano, K.~Ishiwata and N.~Nagata,
Phys.\ Lett.\  B {\bf 690}, 311 (2010)
[arXiv:1004.4090 [hep-ph]].

\bibitem{Sommerfeld}
J.~Hisano, S.~Matsumoto and M.~M.~Nojiri,
Phys.\ Rev.\ Lett.\  {\bf 92}, 031303 (2004)
[arXiv:hep-ph/0307216];
J.~Hisano, S.~Matsumoto, M.~M.~Nojiri and O.~Saito,
Phys.\ Rev.\  D {\bf 71}, 063528 (2005)
[arXiv:hep-ph/0412403];
J.~Hisano, S.~Matsumoto, O.~Saito and M.~Senami,
Phys.\ Rev.\  D {\bf 73}, 055004 (2006)
[arXiv:hep-ph/0511118].

\bibitem{Ackermann:2011wa}
M.~Ackermann {\it et al.}  [Fermi-LAT collaboration],
Phys.\ Rev.\ Lett.\  {\bf 107}, 241302 (2011)
[arXiv:1108.3546 [astro-ph.HE]].

\bibitem{Hryczuk:2011vi}
A.~Hryczuk and R.~Iengo,
arXiv:1111.2916 [hep-ph].

\bibitem{Charbonnier:2011ft}
A.~Charbonnier {\it et al.},
Mon.\ Not.\ Roy.\ Astron.\ Soc.\  {\bf 418}, 1526 (2011)
[arXiv:1104.0412 [astro-ph.HE]].

\bibitem{Adriani:2010rc}
O.~Adriani {\it et al.}  [PAMELA Collaboration],
Phys.\ Rev.\ Lett.\  {\bf 105}, 121101 (2010)
[arXiv:1007.0821 [astro-ph.HE]].

\bibitem{Evoli:2011id}
C.~Evoli, I.~Cholis, D.~Grasso, L.~Maccione and P.~Ullio,
arXiv:1108.0664 [astro-ph.HE].

\bibitem{AMS-02}
\verb$http://www.ams02.org/$.

\bibitem{Adriani:2008zr}
O.~Adriani {\it et al.}  [PAMELA Collaboration],
Nature {\bf 458}, 607 (2009)
[arXiv:0810.4995 [astro-ph]].

\bibitem{PAMELA WINO}
P.~Grajek, G.~Kane, D.~Phalen, A.~Pierce and S.~Watson,
Phys.\ Rev.\  D {\bf 79}, 043506 (2009)
[arXiv:0812.4555 [hep-ph]];
G.~Kane, R.~Lu and S.~Watson,
Phys.\ Lett.\  B {\bf 681}, 151 (2009)
[arXiv:0906.4765 [astro-ph.HE]].

\bibitem{Radiative gluino decay}
M.~Toharia and J.~D.~Wells,
JHEP {\bf 0602}, 015 (2006)
[arXiv:hep-ph/0503175];
P.~Gambino, G.~F.~Giudice and P.~Slavich,
Nucl.\ Phys.\  B {\bf 726}, 35 (2005)
[arXiv:hep-ph/0506214].

\bibitem{ATLAS-CONF}
ATLAS Collaboration,
ATLAS-CONF-2011-155.

\bibitem{Asai:2007sw}
S.~Asai, T.~Moroi, K.~Nishihara and T.~T.~Yanagida,
Phys.\ Lett.\  B {\bf 653}, 81 (2007)
[arXiv:0705.3086 [hep-ph]].

\bibitem{ISR}
J.~Alwall, K.~Hiramatsu, M.~M.~Nojiri and Y.~Shimizu,
Phys.\ Rev.\ Lett.\  {\bf 103}, 151802 (2009)
[arXiv:0905.1201 [hep-ph]];
M.~M.~Nojiri and K.~Sakurai,
Phys.\ Rev.\  D {\bf 82}, 115026 (2010)
[arXiv:1008.1813 [hep-ph]].

\bibitem{Barr:2003rg}
A.~Barr, C.~Lester and P.~Stephens,
J.\ Phys.\ G {\bf 29}, 2343 (2003)
[arXiv:hep-ph/0304226].

\bibitem{Asai:2008sk}
S.~Asai, T.~Moroi and T.~T.~Yanagida,
Phys.\ Lett.\  B {\bf 664}, 185 (2008)
[arXiv:0802.3725 [hep-ph]].


\bibitem{HL-LHC}
\verb$http://hilumilhc.web.cern.ch/HiLumiLHC/$.

\bibitem{ILC}
\verb$http://www.linearcollider.org/$.

\bibitem{CLIC}
\verb$http://clic-study.org/$.

\bibitem{Asano:2011aj}
M.~Asano {\it et al.},
Phys.\ Rev.\  D {\bf 84}, 115003 (2011)
[arXiv:1106.1932 [hep-ph]].


\end{thebibliography}
\end{document}